\documentclass[prl,twocolumn,showpacs,amsmath,amssymb]{revtex4}

\begin{document}

\title{Dirac monopoles embedded into
$SU(N)$ - gauge theory with $\theta$ - term}

\author{M. A. Zubkov}
 \email{zubkov@itep.ru}
\affiliation{ITEP, B.Cheremushkinskaya 25, Moscow, 117259, Russia}

\date{October 1, 2002}

\begin{abstract}
We consider Dirac monopoles embedded into $SU(N)$ gauge theory
with theta-term for
 $\theta = 4\pi M $ (where $M$ is half-integer for $N = 2$ and is integer for
$N>2$).  Due to the theta - term those monopoles obtain the
$SU(N)$ charge and become the dyons. They belong to different
irreducible representations of $SU(N)$ (but not to all of them).
The admitted representations are enumerated. Their minimal rank
increases with increasing of $N$. The main result of the paper is
the representation of the partition function of $SU(N)$ model with
theta-term (that contains singular gauge fields correspondent to
the mentioned monopoles) as the vacuum average of the product of
Wilson loops (considered along the monopole worldlines). This
vacuum average should be calculated within the correspondent model
without theta-term.
\end{abstract}


\maketitle

The concept of monopole (see.~\cite{Dirac}) was originally
associated with singular field configurations in the
electrodynamics, which carry  magnetic charge. Self energy of
Dirac monopoles in the $U(1)$ gauge theory is divergent, which
complicate considerably their physical interpretation.

In nonabelian gauge theories monopoles appear first as solutions
of classical equations of motions in the model that contains
adjoint scalar field ~\cite{Polyakov}. These objects are known as
'tHooft - Polyakov monopoles. The correspondent field
configurations are regular and carry finite energy. In spite of
this 'tHooft - Polyakov monopoles possess some of the
characteristic features of Dirac monopoles, which are connected in
general with quantization of magnetic charge.

Later similar constructions were found within a greate number of
models (see., for example ~\cite{Mon}). The quantum objects which
correspond to these solutions of equations of motion are the
topological defects,i.e. topologically nontrivial field
configurations. In the particular case of the topological defect,
which position is constant in time, the solution of classical
equations of motion (considered within the space-time, from which
the worldline of topological defect is cutted)  reproduces the
classical monopole solution. In such a way Dirac monopoles arise
from the topological defects of $U(1)$ gauge theory ~\cite{Polyak}
while t'Hooft - Polyakov monopoles arise from the topological
defects of Georgy - Glashow model. Below both solution of
classical equation of motion and correspondent topological defect
will be referred to as "monopole". Let us notice that two -
dimensional objects can be considered in analogous way. For
example, Abrikosov - Nielsen - Olesen strings correspond to
quantum topological defects of Abelian Higgs Model ~\cite{forms}.

It follows from the topological consideration that in pure $SU(N)$
gauge theory there is no classical monopole solution. However, due
to the investigation of confinement mechanism in the abelian
projection of gluodynamics and also due to the investigation of
the electroweak theory at finite temperature recently the interest
arises to the topological defects of other models which are
embedded into nonabelian gauge theory ~\cite{B}.
 It is worth mentioning that contrary to the usual monopoles the embedded ones
 are unstable in most cases ( ~\cite{B}, ~\cite{bn}). In spite of this their
 connection with
 the dynamics is rather essential. For example, in the lattice electroweak theory
 at finite temperature their behavior is connected with the nature of
 the Electroweak phase transition
~\cite{C}. In the abelian projection of the gluodynamics the
monopoles correspondent to the remaining $U(1)^{N-1}$ symmetry
~\cite{H} turn out to be responsible for the confinement
~\cite{Pol}. It is important here to note that there exist
different ways to embed abelian monopoles into nonabelian model.
The way we extract abelian variables from nonabelian ones is the
essence of this difference.

More than 20 years ago E.Witten showed that 't Hooft - Polyakov
monopoles become dyons in the model which action contains theta -
term. In ~\cite{Witten} the consideration was of quasiclassical
nature and was applied to the model that contains additional
adjoint scalar field.

In the present work we investigate the influence of the theta -
term on the dynamics of quantum monopoles in the pure  $SU(N)$
gauge theory. We come back to Dirac construction and consider it's
direct generalization to the case of $SU(N)$ gauge theory. The
objects thus arising turn out to be Dirac monopoles, which are
embedded into $SU(N)$ gauge theory and correspond to the factors
of $U(1)^{N-1}$. We omit the question about their possible
instability and about the divergence of the self-energy. (Some
related notes the reader can find at the end of this paper.)

We prove that the monopoles defined in such a way turn out to be
charged and belong to different irreducible representations of
$SU(N)$. Interesting property of the construction is that the
theta - term can be seen only through the appearance of the
$SU(N)$ charge of the monopole. It's topological part proportional
to the number of instantons disappear from the expression of the
partition function for those values of $\theta$, which provide the
monopole with integer $SU(N)$ charge. More explicit, the partition
function of the theory (that belongs to wide class of $SU(N)$
models) with theta-term, which contains singular configurations
correspondent to our monopoles (and does not contain other
singular configurations), is equal to the vacuum average within
the theory without theta-term of the product of the Wilson loops,
which correspond to monopole worldlines. Each wilson loop is
considered in the irreducible representations of $SU(N)$ group,
which is defined by the type of field singularity along the
correspondent monopole worldline.

Let us mention that the definition of the embedded monopole used
in this work differs from the definition of the monopole with even
$Q_m$ considered in ~\cite{PZ}. The last one can be treated as
appeared as a result of application of some singular gauge
transformation to regular gauge field. As it will be seen below
the field singularities which we consider in this paper can not be
constructed in such a way.

Dirac monopoles in four dimensional Euclidian $U(1)$ - gauge
theory are defined as follows ~\cite{Polyak}. Let $A_i$ be the
gauge field and $F_{ij}$ be the correspondent field strength. It
is admitted that $A$ can be singular along the surface  $\Sigma$,
which boundary is the monopole worldline. The field strength is
defined on base of the following expression:
\begin{equation}
i\int_{s} F_{\mu \nu}(y)dx^{\mu} dx^{\nu} = {\rm exp}
(i\int_{\partial s} A_i dx_i) - 1 \label{curvA}
\end{equation}
Here $s$ is infinitely small surface and the parallel transporter
is defined along the boundary of this surface ($y \in
\partial s$). For regular $A$, $ F_{ij}  =
\partial_{[i}A_{j]} $, while along $\Sigma$ this expression is deformed by
subtracting the singularity correspondent to the Dirac string. In
some gauge the field, which corresponds to the monopole worldline
$j$ and it's Dirac string $\Sigma$, is expressed as $A^i =
\bar{A}^i + A_{s}^i[{\Sigma}]$, where $\bar{A}$ is regular part
and $A_s$ satisfies the equation:
\begin{eqnarray}
&&\partial^{[i}A_s^{j]} = 2\pi (^*\Omega)^{ij} +
D^{ij},\nonumber\\
&& \partial^i (^*D)^{ik} = - 2\pi{\cal J}^k\label{AsA}
\end{eqnarray}
Here $\Omega^{ij}(x) = \int_\Sigma \epsilon_{\alpha
\beta}\frac{\partial z^i(\tau)}{\partial \tau_{\alpha}}
\frac{z^j(\sigma)}{\partial \tau_{\beta}}
\delta(x-z(\tau))d^2\tau$, the integration is performed over the
surface $\Sigma$. Points $z$ of this surface are parameterized by
the variable $\tau_{\alpha}, \alpha = 1,2$: $z = z(\tau)$. We
assume that $D$ is regular everywhere except monopole worldline;
$^*D^{ik} = \frac{1}{2}\epsilon^{iklm} D^{lm}$; ${\cal J}^k = -
\int_j \frac{dz^k}{ds} \delta(x-z(s))d s $ (We parameterize points
$z(s)$ of the monopole worldline $j$ by the variable $s$.) Further
it will be usefull to choose $D$ and $A_s$ in such a way, that
near monopole worldline $A_s \sim \frac{1}{r}$, where $r$ is the
distance between the given point and $j$. Also we shall imply that
$A_s^i \frac{dx^i}{ds} = 0$ on the monopole worldline.

We can express the field strength as $ F^{ij} =
\partial^{[i}A^{j]}
 - \pi \epsilon^{ijkl} \Omega^{kl}$. The singularity $\partial_{[i}A_{j]}$ along
  $\Sigma$
is cancelled by the term that contains $\Omega$. Thus the field
strength is singular only along the boundary of this surface.

These singular gauge fields satisfy the equation
\begin{equation}
 \partial^i(^*F)^{ik} =  2\pi {\cal J}^k \label{Bianch}
\end{equation}
Here $(^*F)^{ik} = \frac{1}{2}\epsilon^{iklm} F^{lm}$. Expression
(\ref{Bianch}) shows that our singular field configurations do
represent Dirac monopoles because $\cal J$ is the monopole current
while $^*F$ is the tensor dual to the field strength tensor.

We start to  generalize this construction to the case of  $SU(N)$
theory with the attempt to generalize the equation (\ref{Bianch}).
We consider $SU(N)$ gauge theory in the $4$ - dimensional
Euclidian space and denote the gauge field as $ A_i = A_i^b T_b
\in su(N) $. Here $T_b$ ($b = 1, ..., N^2-1$) are the generators
of  $su(N)$ algebra. This generators are normalized by the
condition ${\rm Tr} \, T_b^2 = 1$. For regular  $A$ the field
strength $ R_{ij} $ is equal to $\partial_{[i}A_{j]} + i
[A_i,A_j]$. The definition of the field strength for the field,
which contain the singularity along  2 - dimensional surface, is
based analogously to the abelian case on the expression $i\int_{s}
R_{\mu \nu}(y)dx^{\mu} dx^{\nu} = {\rm P} \, {\rm exp}
(i\int_{\partial s} A_i dx_i) - 1$. Here again $s$ is the
infinitely small surface and the parallel transporter is defined
along it's boundary.

For the case of regular fields there takes place Bianchy identity
$\partial^i(^*R)^{ik} + i [A^i,(^*R)^{ik}] = 0$, in which
$(^*R)^{ik} = \frac{1}{2}\epsilon^{iklm} R^{lm}$. This identity is
analogous to the abelian identity $\partial^i(^*F)^{ik} = 0$. It
can be deformed to the type that contains monopole current in the
case when the field is singular and the definition of the field
strength is deformed in the correspondent way.

Thus our aim is to found such  configurations of the field $A$,
that their strength satisfy the anomalous Bianchy identity
\begin{equation}
 \partial^i(^*R)^{ik} + i [A^i,(^*R)^{ik}]  =  2\pi {\cal J}^k {\bf n} \label{Bianchy}
\end{equation}
Here matrix ${\bf n}$ is the element of $su(N)$ algebra. Gauge
transformation  $g$ acts on it in the following way: ${\bf n}
\rightarrow g{\bf n}g^+ $ If $j$ has no selfintersections we can
always choose the gauge in which ${\bf n}$ is cartan element. Thus
the abelian nature of the monopoles defined by the condition
(\ref{Bianchy}) is discovered. They must correspond to the factors
of $U(1)^{N-1}$ subgroup, which generators are Cartan elements of
$su(N)$. Such monopoles are Dirac ones embedded into $SU(N)$ gauge
theory.

Below we point out the construction of these objects. As for the
case of $U(1)$ monopoles they appear if we suppose that gauge
field is singular along the surface $\Sigma$, while the field
strength $R$ is admitted to be singular only along it's boundary
$j$ (which is the monopole worldline).
 Both  $j$ and $\Sigma$ are supposed to be smooth and having no
 selfintersections.

Let $S_1$ be the small circle of radius $r$ that links $\Sigma$.
It belongs to the plane orthogonal to $\Sigma$. This plane
intersects $\Sigma$ in the center of the circle. Via the gauge
transformation the fields along all such loops can be made
diagonal.
\begin{eqnarray}
&& A^i s^i = {\bf A}{\bf n} \nonumber\\
&& {\bf n} \in su(N); \, {\bf A} \in R \label{n}
\end{eqnarray}
where ${\bf A}$ and ${\bf n}$ are independent of the point of
$S_1$; $s^i$ is the unity vector which is along $S_1$. The
regularity of the field strength on $\Sigma$ causes the equation
${\rm P} \, {\rm exp}(i\int_{S1} A^i dx^i) = 1  $. The field $A$
satisfies it if we choose
\begin{equation}
{\bf A} \sim  \frac{1}{r} \label{SingS}
\end{equation}
at $r \rightarrow 0$. Diagonal elements of the matrix $\bf n$
should be integer numbers. We consider such  field configurations
that the field in other directions remains regular when we
approach $\Sigma$. Generally the surface $\Sigma$ consists of
several pieces. Each piece of the surface carries it's own $\bf
n$. We can write symbolically ${\bf \Sigma} = \sum_{{\bf n}}\,
\Sigma_{\bf n}$.

Now let us consider the plane (ij) such that the direction (i) is
along $S_1$ and the direction (j) is along $\Sigma$. Then the
requirement that the field strength $R$ should be regular leads
to:
\begin{equation}
[A_k,{\bf n}] = 0,\label{cmt}
\end{equation}
on $\Sigma$. The equation (\ref{cmt}) shows that the gauge field
is effectively abelian along the surface $\Sigma$. Further we
shall imply that (\ref{cmt}) is valid also on the boundary of
$\Sigma$. This additional requirement is rather strong. It
provides that the commutator of the gauge fields cannot contain
the singularity, which would cancel the monopole singularity of
the abelian part of the field strength.

We arrive at the definition of the singular gauge field
correspondent to $U(1)^{N-1}$ - topological defect embedded into
$SU(N)$ theory. Analogously to the $U(1)$ case, this field can be
represented in the following way:
\begin{equation}
A^i = \bar{A}^i + \sum_{\bf n} A_{s}^i[{\Sigma_{\bf n}}] {\bf n}
\label{Reg}
\end{equation}
We suppose here that the gauge is fixed, in which (\ref{n}) takes
place. $\bar{A}$ - is regular part of the gauge field while
$A_s(\Sigma_{\bf n})$ has the form (\ref{AsA}).

The considered monopoles are defined by the formula (\ref{Reg})
and the additional condition (\ref{cmt}) in the gauge in which the
gauge fields are diagonal along all infinitely small circles that
link $\Sigma$. Direct check shows that such a configurations
cannot be constructed from the regular fields by applying any
singular gauge transformations.

Let us take into account the condition (\ref{cmt}). We can express
the field strength that  corresponds to such a field configuration
in the following way:
\begin{equation}
R^{ij} = \partial^{[i}A^{j]} + i [A^i,A^j]
 - \pi \epsilon^{ijkl} \Omega^{kl} {\bf n}
\end{equation}

The singularity of  $\partial_{[i}A_{j]} + i [A_i,A_j]$ along
$\Sigma$ is
 canceled by the term containing $\Omega$.
Thus the field strength is singular only along the boundary of the
surface. It is easy to show that the singular gauge fields of the
form (\ref{Reg}) do possess the anomalous Bianchi identity
(\ref{Bianchy}).

Now let us consider the influence of theta - term on the dynamics
of the constructed monopoles. We have
\begin{eqnarray}
Q & = & \frac{1}{32\pi^2} \int d^4x
 \epsilon_{\mu \nu \rho \sigma} {\rm Tr} R_{\mu \nu} R_{\rho \sigma}\nonumber\\
  & = & \frac{1}{16\pi^2} \int d^4x {\rm Tr}\, G^*G -  \frac{1}{4\pi}\int_{\Sigma} d^2\tau {\rm Tr}\, G_{ij} {\bf n} t_{ij}
\end{eqnarray}
Here $ G_{ij} = \partial_{[i}A_{j]} + i [A_i,A_j]$, ànd $t^{ij} =
\epsilon_{\alpha \beta}\frac{\partial z^i(\tau)}{\partial
\tau_{\alpha}} \frac{z^j(\sigma)}{\partial \tau_{\beta}}$. Using
(\ref{cmt}), we obtain  ${\rm Tr} \, [A_i,A_j] {\bf n} = 0$. Thus
\begin{equation}
Q   =  \frac{1}{16\pi^2} \int d^4x {\rm Tr}\, G^*G -
\frac{1}{4\pi}\int_{j} d x_i {\rm Tr}\, A_{i} {\bf n} \label{ex}
\end{equation}

Careful consideration of the regularization of the  first term in
(\ref{ex}) gives us the evidence that it depends only upon the
values of $A$ at the infinitely distant points. If we suppose that
the field strength should vanish there, we can conclude that this
term appears to be the integer number of instantons $N_{ins}$.
Under this assumption we obtain:
\begin{equation}
{\rm exp}(4\pi M i Q) =
 {\rm exp}(-i\sum_{n} \int_{j_{{\bf n}}}
d x_i M {\rm Tr}\, A^{g[A,\Sigma]}_{i} {\bf n})\label{T1}
\end{equation}
for integer or half-integer $M$. Here $g[A,\Sigma]$ transforms $A$
into the form (\ref{n}) along $\Sigma$.

We consider arbitrary $SU(N)$ model and add the constructed above
monopoles into it's vacuum "by hands". This means that it's
measure $DA_{\Sigma}$ is defined in such a way that it can be
represented as
\begin{equation}
DA_{\Sigma} = D\bar{A} Dg \Delta_{FP}({\bf n})\,
\delta_{\Sigma}({\bf n} - \frac{({\rm Tr}\,{\bf n}^2) \bar{G}^{ij}
(^*t)^{ij}}{{\rm Tr}\,(\bar{G}^{ij} (^*t)^{ij})^2})
\end{equation}
where we represent $A$ as $A = \bar{A}^{g} + \sum_{\bf
n}A_s[\Sigma]g^+{\bf n}g$ and denote $\bar{G}_{ij} =
\partial_{[i}\bar{A}_{j]} + i [\bar{A}_i,\bar{A}_j]$.
 The $\delta$ - function provides the constraint, which is imposed on those components of
$\bar{A}$, which are dual to $\Sigma$. $\Delta_{FP}$ is the
Faddeev - Popov determinant correspondent to this $\delta$ -
function: $\Delta_{FP}^{-1}({\bf n}) = \int D\bar A
\delta_{\Sigma}({\bf n} - \frac{({\rm Tr}\,{\bf n}^2) \bar{G}^{ij}
(^*t)^{ij}}{{\rm Tr}\,(\bar{G}^{ij} (^*t)^{ij})^2})$. Second
constraint $[\bar{A}^i,{\bf n}] = 0$ is supposed to arise
dynamically. We assume that the action $S$ makes field strength as
regular as possible. This provides that it is singular only on the
monopole worldline (not on the string worldsheet). This leads to
$[\bar{A}^i,{\bf n}] = 0$ on $\Sigma$.

The field strength is
\begin{eqnarray}
&&R^{ij}[A] = R^{ij}[\bar{A},g] = g^+ (\bar{G}^{ij}  + 2 \pi {\bf
n} ^*\Omega^{ij}\nonumber\\&& + \sum_{\bf n} (i[\bar{A}^i,{\bf n}]
A_s^j[\Sigma_{\bf n}] + i[{\bf n},\bar{A}^j]A_s^i[\Sigma_{\bf
n}])) g
\end{eqnarray}

The partition function of the model with gauge invariant action
$S[R]$ and with added monopole singularities in the presence of
the theta - term can be expressed as follows:
\begin{eqnarray}
Z_{\Sigma} & = &  \int D A_{\Sigma} exp(-S[R] + 4\pi M i Q[R])\nonumber\\
 & = & \int D \bar{A} D g \Delta_{FP}({\bf n})\,\nonumber\\&& \delta({\bf
n} - \frac{({\rm Tr}\,{\bf n}^2) \bar{G}^{ij} (^*t)^{ij}}{{\rm
Tr}\,(\bar{G}^{ij} (^*t)^{ij})^2})
 {\rm exp}(-S[R[\bar{A},g]])\nonumber\\&&
 \Pi_{\bf n}{\rm exp}(- i \int_{j_{\bf n}} d x_i M ({\rm Tr}\,
(\bar{A}_i {\bf n}))
\end{eqnarray}

Let us consider the following transformation: $\bar{A} \rightarrow
\tilde{A}$, which is defined within small vicinity of $j$. Those
components of $A$, which are along $j$, transform as $\bar{A}
\rightarrow \tilde{A} = h^+ \bar{A} h + h^+ \partial h$ while
other components remain the same: $ \tilde{A} = \bar{A}$. We imply
that $h \rightarrow 1$ on the boundary of mentioned vicinity. The
measure $D\bar{A}$ is obviously invariant under this
transformation. As for the action $S[\bar{A}]$, for the models we
consider in this paper it transforms as $S[\bar{A}] = S[\tilde{A}]
+ O(b)$ (where $b$ is the size of mentioned vicinity). It is due
to the following implied properties of $S$. To be more explicit,
let us first consider $S = \beta \int d^4 x {\rm Tr}\,R^2$. If we
change $\bar{A}$ into $\tilde{A}$, condition $[\tilde{A},{\bf n}]
= 0$ is broken on $j$. Thus $R^{ik} \sim \frac{1}{r}$ (where $(i)$
is along $j$ and $(k)$ is dual to $\Sigma$). Change of action  is
$\Delta S \sim \int d^3 x \frac{1}{r^2} \sim O(b)$. It is worth
mentioning that it is impossible to make such a change $\bar{A}
\rightarrow \tilde{A}$ near all $\Sigma$ with negligible change of
action: in that case constraint $[\bar{A},{\bf n}] = 0$ becomes
broken on all $\Sigma$ and $\Delta S \sim \int d^2 x
\frac{1}{r^2}$ diverges as $\sim {\rm log} \, M$, where $M$ is
ultraviolet cut - off. In this paper we suppose that the action
behaves under the considered transformation in the same way as the
simplest one $S = \beta \int d^4 x {\rm Tr}\,R^2$. Taking into
account this property of the model we have:

\begin{eqnarray}
Z_{\Sigma} & = & {\rm lim}_{b \rightarrow 0}\int D \tilde{A} D h
\Delta_{FP}({\bf n})\,\nonumber\\&& \delta({\bf n} - \frac{({\rm
Tr}\,{\bf n}^2) \tilde{G}^{ij} (^*t)^{ij}}{{\rm
Tr}\,(\tilde{G}^{ij} (^*t)^{ij})^2})
 {\rm exp}(-S[R[\tilde{A},1]] + O(b))\nonumber\\&&
 \Pi_{\bf n}{\rm exp}(- i \int_{j_{\bf n}} d x_i M ({\rm Tr}\,
(\tilde{A^h}_i {\bf n}) )
\nonumber\\
& = & \int D \tilde{A} Dg \Delta_{FP}({\bf n})\,\nonumber\\&&
\delta({\bf n} - \frac{({\rm Tr}\,{\bf n}^2) \tilde{G}^{ij}
(^*t)^{ij}}{{\rm Tr}\,(\tilde{G}^{ij} (^*t)^{ij})^2})
 {\rm exp}(-S[R[\tilde{A},1]])\nonumber\\&&
 \Pi_{\bf n}\, \int \,Dh{\rm exp}(- i \int_{j_{\bf n}} d x_i M ({\rm Tr}\,
(\tilde{A^h}_i {\bf n}) )
\nonumber\\
& = & \int D A_{\Sigma}
 {\rm exp}(-S[R[A]])\nonumber\\&&
 \Pi_{\bf n}\, \int \,Dh{\rm exp}(- i \int_{j_{\bf n}} d x_i M ({\rm Tr}\,
(A^h_i {\bf n}) )
\end{eqnarray}
Here $\tilde{G}_{ij} =
\partial_{[i}\tilde{A}_{j]} + i [\tilde{A}_i,\tilde{A}_j]$.
In the last expression we denote $D A_{\Sigma} = Dg
\Delta_{FP}({\bf n})\, \delta({\bf n} - \frac{({\rm Tr}\,{\bf
n}^2) \tilde{G}^{ij} (^*t)^{ij}}{{\rm Tr}\,(\tilde{G}^{ij}
(^*t)^{ij})^2})$ and $A = \tilde{A}^{g} + \sum_{\bf
n}A_s[\Sigma]g^+{\bf n}g$.

Below we use the abelian representation of the Wilson loop given
in ~\cite{DP}:
\begin{eqnarray}
 {\bf W}_{[{q}]}[j] & = & {\rm Tr} P \, {\rm exp}(i \int_{j} d x_i A_i) \nonumber\\
 & = &\int D \mu_j(g)  {\rm exp}(i \int_{j} d x_i {\rm Tr}\,[A^g]_i {\cal
 H}^{[{q}]})\label{DPW}
\end{eqnarray}
Here the Wilson loop ${\bf W}_{{q}}[j]$ is considered in the
irreducible representation of $SU(N)$ group. The space of this
representation consists of the tensors $\Psi_{i_1 i_2 ... i_r}$.
The symmetry of $\Psi$ is defined by the set of integer numbers
$q_i \, (i = 1, ..., N-1)$ ($\sum_i q_i = r, \, q_i\geq 0 $).
${\cal H}^q = \sum_i {m_i} {\bf H_i}$, where numbers $m_i$
represent the highest weight of the representation while ${\bf
H_i}$ - is the basis of cartan subalgebra of $su(N)$.

Explicit definition of the measure $D\mu_j(g)$ can be formulated
only using some regularization scheme. If we consider lattice
regularization of the model, (\ref{DPW}) can be obtained if we
choose $D\mu_j(g)=(dg_1 D(q))(dg_2 D(q))...(dg_N D(q))$, where
$\int dg = 1$ and $D(q)$ is the dimension of representation. (This
definition follows from completness identity for coherent system:
$\int dg |g\rangle \langle g| = \frac{1}{D(q)}$.) This means that
in continuum theory (\ref{DPW}) should be used with careful
consideration of ultraviolet divergences. Particulary, integration
measure can be expressed through the conventional measure $Dg$
(such that $\int Dg = 1$) as follows:
\begin{equation}
D\mu_j(g) = {\rm exp}( m_q |j|) Dg,
\end{equation}
where $|j|$ is the length of the Wilson loop while $m_q$ diverges
as $m_q \sim M \,{\rm log}\,D(q)$, where $M = \frac{1}{a}$ is
ultraviolet cut-off and $a$ is lattice spacing.  For any
particular model $m_q$ must be included into renormalization
scheme for the mass of correspondent particle.

Direct calculation shows that nonzero elements of diagonal matrix
 ${\cal H}^{q}$ are expressed as follows ($q_N = 0$):
\begin{equation}
 {\cal H}^{q}_{ii} = q_i - \frac{1}{N} \sum_k q_k \, (i = 1, ..., N)\label{H}
\end{equation}

For any matrix  ${\bf n} = {\rm diag}(n_1,...,n_N)$ that lives on
the worldline of the monopole we can choose the representation of
$SU(N)$ group as follows. Let us arrange the values of $n_i$ in
such a way that $n_{N} \geq n_{N-1} \geq ... \geq n_1$.
 The correspondent representation of the gauge group is defined by
the following set of $q_i$:
\begin{equation}
q_i({\bf n}) =  M \, (n_{N} - n_{i})
\end{equation}

The chosen representation is denoted as $[q({\bf n})]$. We can
rearrange the diagonal elements of ${\cal H}^{[q({\bf n})]}$ in
such a way that it will become
 equal to the matrix $- M {\bf n}$. This leads to the following
representation of the partition function:
\begin{equation}
   Z_{\Sigma} = \int D A_{\Sigma}  e^{ - S[R[A]] } \nonumber\\
 \Pi_{\bf n} e^{- m_{q({\bf n})} |j_{\bf n}|}{\bf W}_{[q({\bf
n})]}[j_{\bf n}]> ,
\end{equation}
where $j_{\bf n}$ is the worldline of the monopole, which carries
the matrix $\bf n$. As was explained above naively divergent
constant $m_q$ appears, which should be absorbed (in correctely
defined model) by the renormalization of monopole mass. Taking
into account this absorbtion we redefine the action $S[A]$ as $S
\rightarrow \bar{S} = S + \sum_{\bf n} m_{q({\bf n})} |j_{\bf n}|
$. Finally we arrive at the following expression for the partition
function:
\begin{eqnarray}
   Z_{\Sigma}&=& \int D A_{\Sigma} e^{ - \bar{S}[A] }
   \Pi_{\bf n} {\bf W}_{[q({\bf n})]}[j_{\bf n}]\nonumber\\
   &=& \langle \Pi_{\bf n} {\bf W}_{[q({\bf n})]}[j_{\bf n}]\rangle  , \label{ZH}
\end{eqnarray}
In the last expression average is over the model without theta
term and with action $\bar{S}$.

In general case in order to obtain integer numbers of $q$, $M$ is
required to be integer. But for the particular case of $SU(2)$
group  $\sigma_3$ is the only Cartan element. Thus $q = M \, {\rm
Tr}\,({\bf n} \sigma_3)$ is integer for any half-integer $M$. For
this reason we consider in this paper $\theta = 4 \pi M $, where
$M$ is half-integer for $N=2$ and is integer for $N>2$.

The expression (\ref{ZH}) gives the evidence that the monopoles
become dyons. Peculiar feature of this expression is that the
topological term disappears from the partition function. Thus the
only influence of the theta - term on the dynamics is that due to
it the monopoles become charged.

Let us mention that not all of the representations appear in
(\ref{ZH}). To be more explicit let us enumerate the admitted
representations. Those representations are defined by the sets
$(q_1, ..., q_{N-1})$ such that $\sum_i q_i = N M L$, where
 $L$ is integer. For $N = 2$ and $M = \frac{1}{2}$
 the full set of the irreducible representations of $SU(2)$ appears.
 For $N=3$ and $M=1$ the lowest admitted representations are:
$(3, 0), (2,1), (6, 0), (5, 1), (4, 2) ...$. For $SU(5)$ group the
lowest admitted symmetric representation is $(5,0,0,0)$.

This way our consideration of Dirac monopoles embedded into the
$SU(N)$ gauge theory with theta - term shows that those monopoles
become dyons. It is shown that the partition function of the
theory with  monopole singularities added "by hands" is expressed
as the vacuum average of the product of the Wilson loops that
correspond to the monopoles. The average should be calculated
within the $SU(N)$ theory without theta - term. This expression
contains the infinite number of irreducible representations of the
gauge group. The peculiar feature of the theory is that the usual
topological charge disappears from the expression of the partition
function. Another interesting result is that
 the rank of the representations, which appear in (\ref{ZH}), increases with
increasing of $N$.

As it was mentioned above in the present work we consider monopole
singularities as external and add them "by hands" into the
functional integral. On the present level of understanding we
cannot consider these objects as appeared dynamically because (as
for the case of Dirac monopole) we do not define the model, in
which the correspondent singularity of the gauge field leads to
finite values of the action. Furthermore, the consideration of
~\cite{PZ} shows that such configurations may be unstable.
Nevertheless let us note that a lot of properties of "unreal"
(from the point of view of abelian theories) Dirac monopoles turn
out to be inherent to the objects of finite energy which are
'tHooft - Polyakov monopoles, appeared in more complex model. It
might be possible that in the same way properties of the
considered above Dirac monopoles embedded into $SU(N)$ gauge
theory with theta-term will turn out to be inherent to realistic
objects of some more complex physical model.

\begin{acknowledgments}
    The author is grateful to M.I.Polykarpov,  V.I.Zakharov and F.V.Gubarev for
useful and stimulating discussions. This work was partly supported
by the grants INTAS 00-00111 and CRDF award RP1-2364-MO-02.
\end{acknowledgments}


\begin{thebibliography}{13}
\expandafter\ifx\csname
natexlab\endcsname\relax\def\natexlab#1{#1}\fi
\expandafter\ifx\csname bibnamefont\endcsname\relax
  \def\bibnamefont#1{#1}\fi
\expandafter\ifx\csname bibfnamefont\endcsname\relax
  \def\bibfnamefont#1{#1}\fi
\expandafter\ifx\csname citenamefont\endcsname\relax
  \def\citenamefont#1{#1}\fi
\expandafter\ifx\csname url\endcsname\relax
  \def\url#1{\texttt{#1}}\fi
\expandafter\ifx\csname
urlprefix\endcsname\relax\def\urlprefix{URL }\fi
\providecommand{\bibinfo}[2]{#2}
\providecommand{\eprint}[2][]{\url{#2}}

\bibitem{Dirac}
P.A.M. Dirac, Proc. Roy. Soc. {\bf A 133},  60 (1931).

\bibitem{Polyakov}
A.M. Polyakov, Pis'ma v ZhETF {\bf 20},  430 (1974).



\bibitem{Mon}
"Monopoles. Topological and variational methods." The collection
of articles edited by M.I. Monastyrsky and A.G.Sergeev, Moscow,
"MIR", 1989.

\bibitem{Polyak}
A.M. Polyakov, "Gauge Fields and strings" Harwood Academic
Publishers (1987).

\bibitem{forms}
M.I. Polikarpov, U.J. Wiese, and M.A. Zubkov, Phys. Lett. {\bf B
309} (1993) 133

\bibitem{B}
M. Barriola, T. Vachaspati and M. Bucher, Phys. Rev. {\bf D 50},
2819 (1994).

\bibitem{bn}
R.A. Brandt, F. Neri, Nucl. Phys. {\bf B 161} (1979) 253

\bibitem{C}
M.~N. Chernodub, F.~V. Gubarev, E.~M. Ilgenfritz et all, Phys.
Lett. {\bf B 434}, 83 (1998).

\bibitem{H}
G. 't Hooft, Nucl. Phys. {\bf B 190}, 455 (1981).

\bibitem{Pol}
M.~I. Polikarpov, Nucl. Phys. Proc. Suppl. {\bf 53}, 134 (1997).

\bibitem{Witten}
E. Witten, Phys. Lett. {\bf B 86},  283 (1979).

\bibitem{PZ}
M.~N. Chernodub, F.~V. Gubarev, M.~I. Polikarpov et all,
 Nucl. Phys. {\bf B 592}, 107 (2001).




\bibitem{DP}
D. Diakonov and V. Petrov, Phys. Lett. {\bf B 224}, 131 (1989).








\end{thebibliography}
\end{document}